# A Dynamically Reconfigurable Terahertz Array Antenna for Near-field Imaging Applications

Kathirvel Nallappan[1], Jingwen Li[1], Hichem Guerboukha[1], Andrey Markov[1], Branko Petrov[2], Denis Morris[2] AND Maksim Skorobogatiy[1,*]

[1]Department of Physics Engineering, École Polytechnique de Montréal, Montréal, Québec, Canada
[2]Department of Physics, University de Sherbrooke, Québec, Canada
*Corresponding author: maksim.skorobogatiy@polymtl.ca

**A proof of concept for high speed near-field imaging with sub-wavelength resolution using SLM is presented. An 8 channel THz detector array antenna with an electrode gap of 100 μm and length of 5 mm is fabricated using the commercially available GaAs semiconductor substrate. Each array antenna can be excited simultaneously by spatially reconfiguring the optical probe beam and the THz electric field can be recorded using 8 channel lock-in amplifiers. By scanning the probe beam along the length of the array antenna, a 2D image can be obtained with amplitude, phase and frequency information.** © 2017 Optical Society of America

Terahertz (THz) frequency ranges from 0.1-10 THz with their corresponding wavelengths spanning from 3 mm to 30 μm in the electromagnetic spectrum. With many interesting properties such as non-ionizing, transparency to wide range of dry materials, molecular fingerprinting etc., THz imaging and diagnostics have a tremendous potential for applications in non-destructive testing and imaging [1-4], medical diagnosis [5-6], health monitoring [7] and chemical and biological identification [8-10].

For applications in the THz imaging a higher number of electric field traces have to be taken as one has to resolve two lateral dimensions and one time dimension. A typical single-pixel THz pulse imaging system therefore requires scan times ranging in hours. The use of special scanning technique such as circular delay stage [11] can decrease the scanning time associated with a time axis, however the problem of a spatial 2D scanning of a sample still remains. Other methods try to accelerate the spatial acquisition by forging the spatial scanning using 2D electro-optic sampling with a CCD camera [12]. However, due to the lack of the real lock-in techniques for CCD cameras, expensive low repetition (~1 KHz) laser amplifier systems with high femto-second (fs) pulse energies are required. As a consequence, the signal to noise ratio (SNR) is considerably worse than with single pixel detection by a photoconductive antenna. Another approach was demonstrated by controlling the electromagnetic properties of the semiconductor mask spatially and dynamically to increase the measurement speed [13]. Hairui Liu et al., developed an extended hyper-hemispherical silicon lens with a focal plane array for 1D measurement [14]. Recently, Anika Brahm et al., demonstrated the 1D multichannel imaging system with 15 detection channels with the increase in the measurement speed by a factor of 15 [15]. As the detector antenna is typically sensitive to a single polarization of the electric field, in order to measure E-field vector, the measurements have to be repeated twice for the two detector orientations.

It was long recognized that the multichannel detection using photoconductive antenna arrays [16-17] constitutes a viable detection technique such as arrays promise an outstanding SNR due to the use of lock-in technique. Although the idea of parallel interrogation of antenna arrays is not new, practical realization of this idea lagged behind due to engineering complexity. One of the first successful demonstrations of a multichannel THz detection was accomplished only several years ago [18]. In that work, the THz pulse was detected by a 16 pixel photoconductive antenna array in which the core idea was to split the main fs laser beam into several smaller beams and then focus them onto individual antenna using microlens array. However, the spatial resolution is limited to 0.5 mm due to the difficulty in fabrication and alignment of the microlens array on top of the antenna array.

In this work, we proposed and fabricated the detector array antenna with the capability of both 1D and 2D measurements by dynamically steering the fs beam using the spatial light modulator (SLM). Antenna array with eight dipole structures are fabricated using the Gallium Arsenide (GaAs) semiconductor substrate whose bandgap is suitable for the 800 nm excitation. A rapid thermal annealing at 410°C for 30 S was carried out to bury the metal contacts. The gap between the electrodes is 100 μm and the length of the dipole is 5 mm. The fabricated array antenna [Fig. 1(a)] is packaged in the PCB board [Fig. 1(b)] for further characterization.

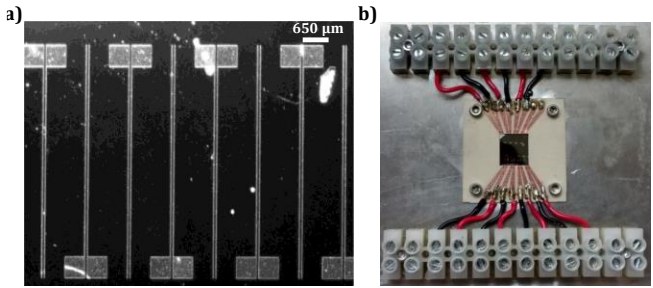

Fig. 1. (a) Fabricated array antenna with 8 dipole element and (b) Packaged in the PCB board for characterization

The response of the fabricated antenna is characterized by measuring the photocurrent and the resistance using two point probe station. The V-I curve as shown in Fig. 2 presents the behavior of one of the antenna when the lamp is ON and OFF.

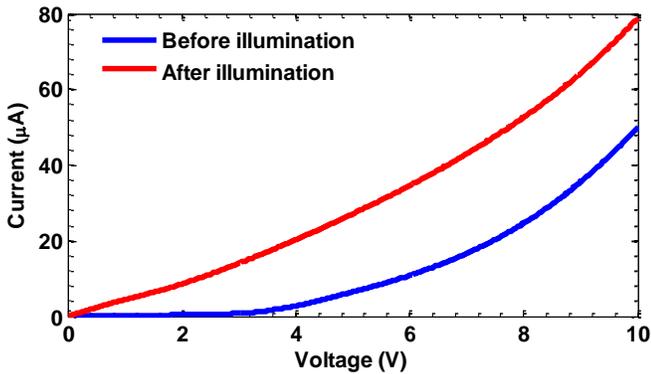

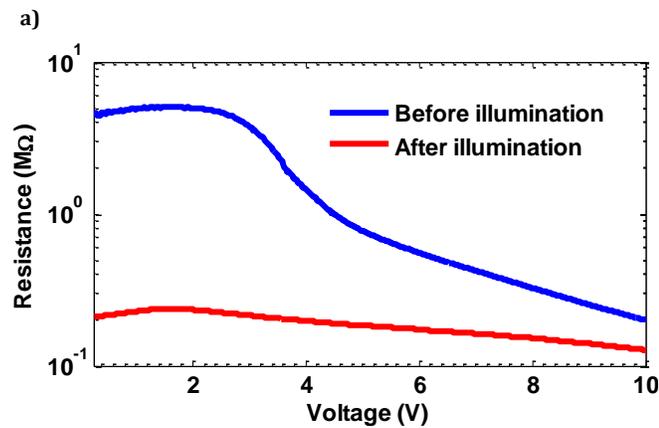

Fig. 2. Response of one of the dipole antenna when the LED lamp is turned ON and OFF (a) V-I characteristics and (b) Resistivity

The dark current and the dark resistance shows the prominent Schottky behavior which could be due to the nature of the substrate which is not being enough resistive. However, the response is linear above 6V which does not affect the performance of the antenna.

Fig. 3. Shows the schematic of the proposed experimental set up for the characterization of the array antenna. A Ti: Sapphire fs laser with the pulse duration of 100 fs, repetition rate of 80 MHZ and ~3W average power from spectra physics (Mai Tai HP) is used as the excitation source. An interdigitated antenna from Batop Optoelectronics (iPCA-21-05-1000-800-h) is used as the THz emitter.

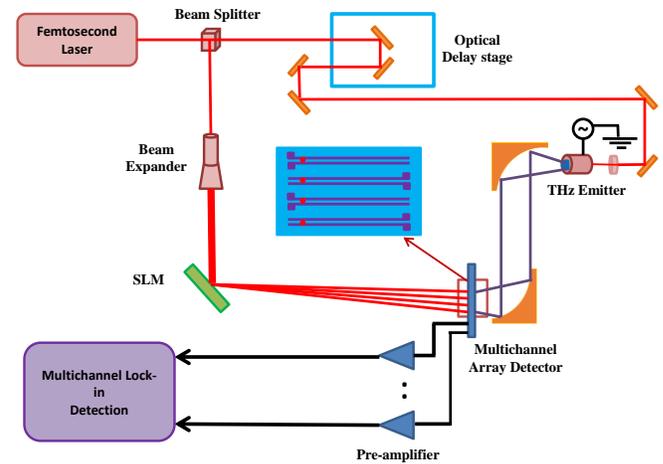

Fig. 3. Schematic of the experimental set up: The phase of the incident optical beam is configured to focus at the gap of each dipole antenna. Each dipole element is individually connected to the lock-in amplifier.

A pump power of 500 mW (average) at the wavelength of 770 nm is focused onto the emitter antenna. A square wave bias modulation with the peak to peak amplitude of 14 Vpp at the frequency of 40 KHz is used for lock-in detection purpose. For the initial settings and measurements, a single pixel photoconductive detector antenna from the Menlo Systems (TERA8-1) is used. The THz path length is aligned carefully to estimate the maximum THz photocurrent from the emitter. By probing the detector with the average power of 10 mW, the THz photocurrent with the peak amplitude of ~50 nA is recorded as shown in Fig. 4.

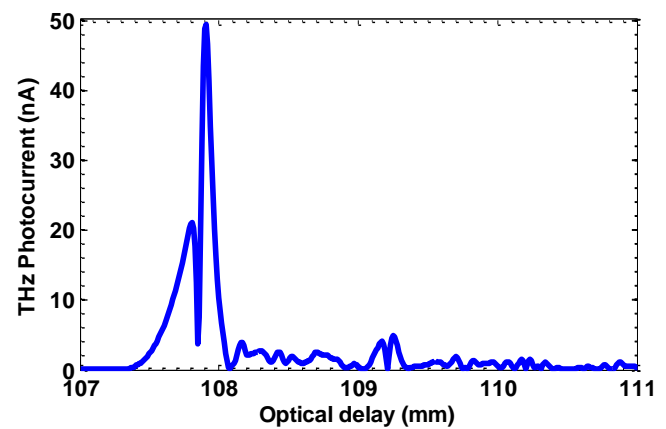

Fig. 4. THz photocurrent recorded from the interdigitated emitter antenna with the average pump power of 500 mW. The average power of 10 mW is used for the detector antenna (TERA8-1).

The SLM from HOLOEYE Photonics is used for modulating the fs light spatially and focus on to the desired location. The SLM

feature a high spatial resolution with an array of 1920 X 1080 pixels. The size of each pixel is 8 µm and has 10 bit-depth encoding to create a reconfigurable phase profile. . It is characterized by focusing the expanded laser beam and then subdivided into several spots at the desired focal plane without any additional optics. To create a desired pattern at the focal plane, we have implemented a classic Gerchberg-Saxton algorithm for the calculation of the phase distribution at the SLM plane. Multiple focal spots are created using the SLM and recorded on the CCD at the focal distance of 50 cm. as shown in Fig. 5. The minimal focal spot size is of the order of 30 µm (Full width Half Maximum) and the spatial accuracy of 5 µm is achieved.

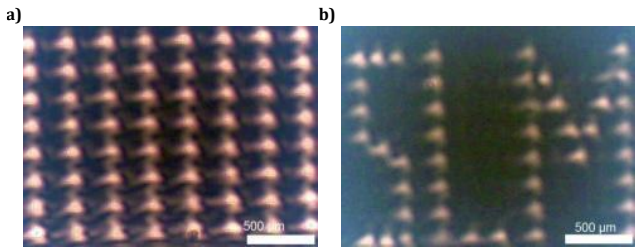

Fig. 5. The phase of the incoming laser beam is configured to create multiple focal spots at a distance of 50 cm (a) linear pattern and (b) alphabets demonstrating angular focal spots.

The basic principle behind the proposed scheme by using SLM is to subdivide the optical beam into an N individual beams and focus them individually on the electrode gap of the N dipole antennas.

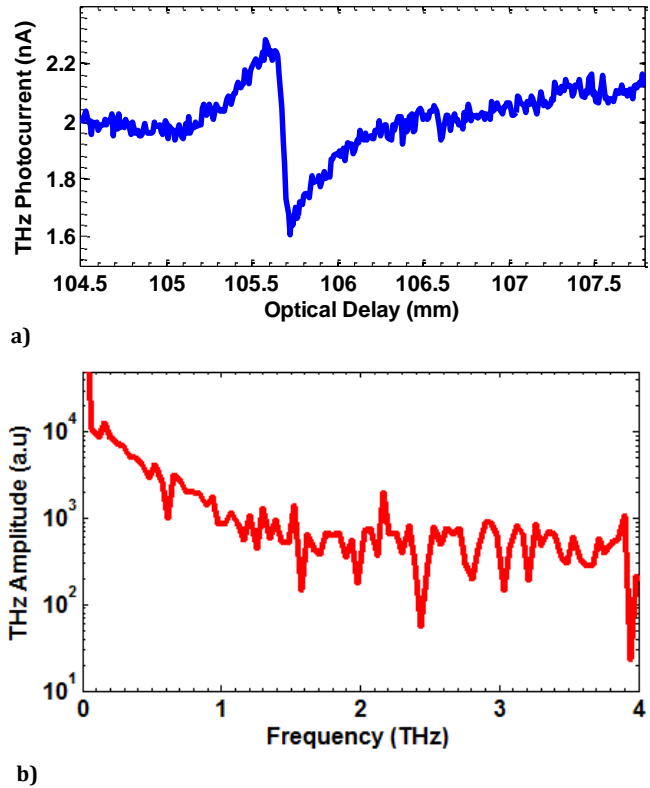

Fig. 6. Characterization of one of the dipoles in the array antenna. (a) Detected THz pulse in time domain and (b) corresponding frequency spectrum.

Additionally, by introducing spatially varying linearly changing phase shifts, the SLM can be used to deflect the array of focused beams to scan parts of the N antenna sequentially (E.g. row by row as shown in Fig. 3.). This approach constitutes a novel versatile interrogation platform as it allows dynamic reconfiguring of the THz detector to explore the trade-off between the acquisition speed and the signal to noise ratio. A fast 1D measurement and relatively slower 2D measurements

The signal pixel detector is replaced by the fabricated array antenna and only one of the dipole is excited by the optical probe beam. Fig. 6(a) presents the THz pulse in the time domain scanned with the optical delay step of 0.01 mm. The corresponding frequency spectrum is shown in Fig. 6(b). However, the THz power of our system is limited to perform the scanning of all the dipole antenna simultaneously.

In conclusion, we have presented a novel design with THz array antenna to acquire high speed near-field THz imaging using SLM. An SLM is used to reconfigure the optical probe beam spatially and dynamically allowing to perform both 1D and 2D measurements with subwavelength resolution. Also, the THz electric field vector can be directly measured using the proposed approach.

**Acknowledgment**. We thank Mr. Traian Antonescu, Mr. Jules Gauthier and Mr. Francis Boutet for their help in the fabrication of PCB board, wire bonding and fabricating the antenna holder.